**Владимиренко Максим Андрійович**
аспірант кафедри інформаційної та кібернетичної безпеки
Київський університет імені Бориса Грінченка, Київ, Україна
OrcID: 0000-0001-8852-7939
*reinekekun@gmail.com*

**Соколов Володимир Юрійович**
старший викладач кафедри інформаційної та кібернетичної безпеки
Київський університет імені Бориса Грінченка, Київ, Україна
OrcID: 0000-0002-9349-7946
*vladimir.y.sokolov@gmail.com*

**Астапеня Володимир Михайлович**
кандидат технічних наук, доцент,
доцент кафедри інформаційної та кібернетичної безпеки
Київський університет імені Бориса Грінченка, Київ, Україна
OrcID: 0000-0003-0124-216X
*v.astapenia@kubg.edu.ua*

# ДОСЛІДЖЕННЯ СТІЙКОСТІ РОБОТИ ОДНОРАНГОВИХ БЕЗПРОВОДОВИХ МЕРЕЖ ІЗ САМООРГАНІЗАЦІЄЮ

**Анотація.** На сьогоднішній день з'явилася необхідність у розробці дієвих протоколів обміну даними та пристроїв, що будуть цей обмін забезпечувати, оскільки стандартні протоколи, що використовуються у традиційних мережах не можуть в повній мірі задовольнити потреб нового типу мереж. У статті описано процес розробки та реалізації натурної моделі стійкої до завад та розривів сенсорної мережі. Стійкість даної мережі досягається шляхом побудови розподіленої мережі, в якій всі вузли передають повідомлення всім доступним вузлам. Безпроводові мобільні однорангові мережі (MANET) можуть автоматично конфігуруватися, тому вузли в ній можуть вільно переміщатися. Безпроводові мережі не мають складнощів налаштування інфраструктури та управління, що дозволяє пристроям створювати і приєднуватися до мереж «на льоту» — в будь-якому місці, в будь-який час. У даній роботі розглянуто теоретичну частину функціонування подібних мереж та галузі їх використання. Після цього проведено початковий аналіз доступного обладнання, що використовується для побудови подібних апаратних рішень. Детально розглянуто програмне забезпечення для розробки таких рішень, а також наведено приклади готових моделей, що реалізують досліджуваний функціонал. Після цього зібрано декілька варіантів натурної моделі мережевих вузлів, а також тестового приладу для створення корисного навантаження на мережу. Для цього були використані сторонні відкриті рішення у поєднанні з власними розробками. З отриманою системою проведено ряд тестів, що дали змогу зрозуміти слабкі й сильні сторони такої мережі та зробити висновки для подальшого розвитку проекту та створення вдосконаленого робочого прототипу. У статті наведено принципові електричні схеми пристроїв, список використаного обладнання та програмного забезпечення, що було використано та наведено фотоматеріали прототипів створеної системи. Дана система може бути використана в реальних умовах для утворення системи розумного дому, отримання інформації з певних IoT датчиків.

**Ключові слова:** сенсорна мережа; мережа із самоорганізацією; обмін даними; мікроконтролер; Wi-Fi; відмовостійкість; IoT; безпроводовий датчик.

## 1. ВСТУП

На сьогоднішній день розвиток мережевих технологій сягає все більших об'ємів, що позначається на кількості цифрових пристроїв, що підключені до мережі. Дане явище, коли кожен пристрій отримав змогу обмінюватися службовими даними, отримало назву «інтернет речей» (internet of things, IoT). При такому швидкому розвитку подій, з'явилася необхідність у розробці дієвих протоколів обміну даними та пристроїв, що будуть цей обмін забезпечувати, оскільки стандартні протоколи, що використовуються у традиційних мережах не можуть в повній мірі задовольнити потреб нового типу мереж, що стали називати сенсорними. Дані мережі повинні мати змогу до самоорганізації, оскільки налаштовувати вручну мережу на кожному новому пристрої досить проблематично. Окрім того, вони повинні автоматично відновлювати розірвані з'єднання та знаходити нові маршрути при динамічній зміні топології мережі, оскільки більшість з пристроїв, які об'єднує концепція IoT є досить мобільними, тому закріплення їх у мережі не уявляється можливим.

Таким чином у даній роботі розглядаються властивості таких мереж, а також способи їх оптимізації задля підвищення рівня відмовостійкості та швидкості обміну даними.

## 2. НАДІЙНІСТЬ ПЕРЕДАЧІ ДАНИХ У БЕЗПРОВОДОВІЙ МЕРЕЖІ

Безпроводова однорангова мережа ad hoc (від лат. «для певного випадку») є децентралізованою безпроводовою мережею. Мережу можна назвати «спеціальною» для певного випадку (відповідно до її назви) бо вона не спирається на вже існуючу інфраструктуру, таку як маршрутизатори в провідних мережах або точки доступу (ТД) в маршрутизованій мережі [1]. Замість цього, кожен вузол бере участь в маршрутизації шляхом пересилання даних іншим вузлам, таким чином, визначення через які вузли будуть передані дані здійснюється динамічно на основі підключення до мережі і алгоритму маршрутизації, що використовується в певному випадку.

Безпроводові мобільні однорангові мережі (mobile ad hoc network, MANET) можуть автоматично конфігуруватися, тому вузли в ній можуть вільно переміщатися. Безпроводові мережі не мають складнощів налаштування інфраструктури та управління, що дозволяє пристроям створювати і приєднуватися до мереж «на льоту» — в будь-якому місці, в будь-який час [2].

Децентралізований характер безпроводових мереж робить їх придатними для різних застосувань, де центральні вузли не можуть бути прокладені, а також може поліпшити масштабованість мереж в порівнянні з безпроводовими керованими мережами [3]. Мінімальна конфігурація і швидке розгортання роблять однорангові мережі підходящими для надзвичайних ситуацій, таких як стихійні лиха або військові конфлікти. Наявність динамічних і адаптивних протоколів маршрутизації дозволяє одноранговим мережам швидко конфігуруватися та видозмінюватися [4].

Головні обмеженням при роботі з вузлами такої мережі, є те, що вони можуть бути рухливими, в результаті чого з'єднання будуть часто перериватися і відновлюватися. Крім того, смуга пропускання безпроводового каналу також обмежена, а вузли працюють на обмеженому живленні від батареї, ресурс якої вичерпний.

Як і в традиційних протоколах маршрутизації, вузли ad hoc мережі можуть працювати з таблицями маршрутизації. Зокрема, використовуються відстань-векторні протоколи, що засновані на обчисленні напрямку і відстані до будь-якої лінії зв'язку в мережі. «Напрямок» зазвичай включає в себе наступну адресу до вузла і інтерфейс виходу. «Відстань» є мірою вартості, для  досягнення певного вузла. Найменша вартість маршруту між будь-якими двома вузлами і маршрут з мінімальною відстанню. Кожен вузол підтримує вектор мінімальної відстані до кожного вузла. Вартість досягнення пункту призначення розраховуються з використанням різних метрик маршруту. Протокол інформаційної маршрутизації (routing information protocol, RIP) використовує лічильник переходів за призначенням, тоді як внутрішньо-шлюзовий протокол (interior gateway routing protocol, IGRP) бере до уваги іншу інформацію, таку як затримка вузла і доступність смуги пропускання.

Що стосується безпеки даних, що передаються сенсорними мережами, варто врахувати, що більшість ad hoc мереж не здійснюють ніякого контролю доступу до мережі, в результаті чого ці мережі уразливі до атак ресурсів споживання, де зловмисний вузол вводить пакети в мережу з метою виснаження ресурсів вузлів ретрансляції пакетів [5].

Для того, щоб перешкодити або запобігти подібні атаки, необхідно використовувати механізми автентифікації, які забезпечують тільки авторизовані вузли можуть вводити трафік в мережу [6]. Проте треба розуміти, що навіть з автентифікацією, ці мережі уразливі для пакетів деавторизації або уповільнюючих атак, в результаті чого проміжний вузол відкидає пакет або затримує його, а не відразу ж відправляє його на наступний вузол.

Таким чином спираючись на залежність принципів роботи мережі від її призначення, необхідно чітко виділити вимоги до мережі, що проектується, перш ніж починати роботу над нею, аби уникнути ситуації, через які в подальшому доведеться вносити правки до цілісної архітектури програми.

## 2. АПАРАТНО-ПРОГРАМНОГО ЗАБЕЗПЕЧЕННЯ СЕНСОРНОЇ МЕРЕЖІ

Для реалізації даного проекту було вирішено використати мікроконтролер (МК) ESP8266, оскільки він пропонує необхідний для подальшої роботи функціонал у вигляді Wi-Fi модулю, що може працювати у b/g/n режимах, що регламентовані у стандарті IEEE 802.11. При цьому даний модуль має низьку ціну порівняно з іншими подібними рішеннями, що у результаті робить його оптимальним рішенням для побудови невеликих, і що головне — недорогих сенсорних мереж.

Даний МК реалізований у декількох варіантах (список модулів в табл. 1), проте найрозповсюдженішими є два з них: ESP-01 та ESP-12E. Саме їх ми і будемо використовувати у проекті, враховуючи їх характеристики та доступність.

У кінцевому проекті планується використовувати ESP-01 через його просту та доступність. Не дивлячись на те, що модуль ESP-01 має певний недолік у вигляді малої кількості інтерфейсів введення/виведення загального призначення (general purpose input/output, GPIO), через є проблемним керування сторонніми пристроями напряму через сигнали, що подаються через дані інтерфейси, проте це не є критичним критерієм при виборі, оскільки головною ціллю пристрою є передача даних, отриманих через послідовний інтерфейс від головного пристрою та передача через послідовний інтерфейс даних, які були отримані по Wi-Fi. Таким чином ціллю ESP-01 є розширення функіоналу вже існуючих систем, а саме додавання можливості комунікації з ними.

*Таблиця 1*

**Модельний ряд модулів на базі МК ESP8266**

| Модуль | Кількість активних GPIO | Розміри, мм | Особливості |
|---|---|---|---|
| ESP-01 | 6 | 14,3×24,8 | Найпоширеніша модель. При ініціалізації модулю задіюються всі GPIO через що використовувати їх у власних потребах проблемно. Не підтримує режим глибокого сну |
| ESP-02 | 6 | 14,2×14,2 | Власна антена відсутня. Є роз'єм для зовнішньої антени |
| ESP-03 | 10 | 17,3×12,1 | Керамічна антена. Виведено всі GPIO |
| ESP-04 | 10 | 14,7×12,1 | Антена відсутня. Виведено усі GPIO |
| ESP-05 | 3 | 14,2×14,2 | Виведені лише основні GPIO: VCC, GND, TX, RX, RST |
| ESP-06 | 11 | 14,2×14,7 | Верхня частина повністю екранована. Усі виводи знизу |
| ESP-07 | 14 | 20,0×16,0 | МК на платі екрановано. Вбудована керамічна антена. Також є роз'єм для підключення зовнішньої |
| ESP-08 | 10 | 17,0×16,0 | Такий самий модуль як і ESP-07, але без вбудованої антени |
| ESP-09 | 10 | 10,0×10,0 | Найменший модуль |
| ESP-10 | 3 | 14,2×10,0 | ESP-09 з деякими виправленими проблемами попередньої моделі |
| ESP-11 | 6 | 17,3×12,1 | Від ESP-10 відрізняється більшою кількістю виведених GPIO та керамічною антеною |
| ESP-12 | 14 | 24,0×16,0 | МК екрановано. Збільшено об'єм динамічної пам'яті |
| ESP-12E | 20 | 24,0×16,0 | На лівій стороні модулю виведено 6 додаткових GPIO |

Якщо ж є необхідність у використанні GPIO — варто використати ESP 12E — більш дорожчу, проте й більш функціональну версію модулю. Даний модуль має 20 GPIO, які ми можемо використовувати на свій розсуд в тій чи іншій мірі. Тому безпосередньо на етапі тестування буде використано також і цей модуль, оскільки керовані GPIO надають нам можливість відтворити повну тестову модель (рис. 1), що складається з «батьківського» пристрою, який виконує певні дії та додаткового модулю (в нашому випадку ESP-01), що передає дані по Wi-Fi.

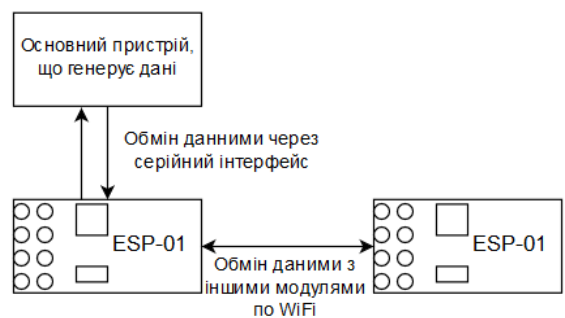

*Рис. 1. Концепція взаємодії модуля ESP-01 з іншими пристроями*

У підключенні модулю ESP-01 є деякі певні нюанси, оскільки даний модуль потребує підключення деяких GPIO до відповідних контактів, так як під час запуску модулю, МК перевіряє рівень напруги на цих GPIO та в залежності від цього виконує певні дії. Найпростіша схема підключення модуля зображена на рис. 2. ESP-01 всього має 8 виведених для підключення GPIO, серед яких:

– VCC — контакт для підключення живлення. Для коректної роботи необхідно підключити до напруги від 3 до 3,8 В;

– GND — заземлення;

– CH_PD — контакт необхідно під'єднати до живлення через підтягуючий резистор на 10 кОм (R1 на рис. 2) для роботи модуля;

– RXD — контакт для отримання даних через послідовний інтерфейс;

– TXD — контакт для надсилання даних через послідовний інтерфейс;

– IO0 — контакт, що контролює режим роботи модуля. Для нормального режиму повинен бути під'єднаний до живлення через підтягуючий резистор на 10 кОм (R3 на рис. 2). При замиканні з землею модуль перейде в режим для завантаження мікропрограми на нього;

– IO2 — даний контакт можна використовувати за потребою, проте слід пам'ятати, що при запуску модулю він використовується мікроконтролером, який подає на нього сигнал. Тому, якщо на контакт буде під'єднаний, наприклад, світлодіод, при запуску він блимне;

– RST — контакт для перезавантаження модуля. Для перезавантаження необхідно замкнути з землею. В інший час май бути під'єднаний до живлення через підтягуючий резистор на 10 кОм (R2 на рис. 2).

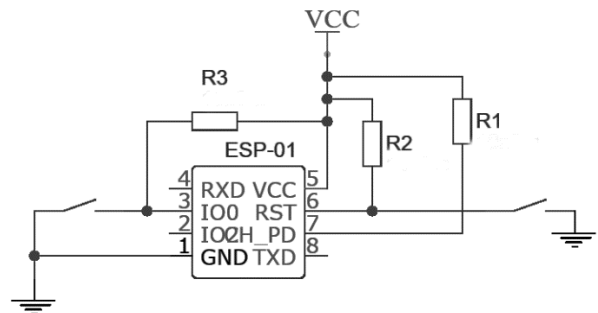

*Рис. 2. Принципова схема підключення модулю ESP-01 для початку роботи*

Після підключення GPIO так як це було зазначено вище, модуль почне працювати, про що свідчитиме запалений червоний світлодіод.

Також на модулі є світлодіод синього кольору, робота якого сигналізує про активність на послідовному інтерфейсі модуля. Для того, аби побачити, які дані надсилає модуль, або ж самостійно надіслати на якийсь з модулів — необхідно використати конвертер для переведення сигналу з універсального асинхронного прийомопередатчика (universal asynchronous receiver-transmitter, UART) та універсальну послідовну шину (universal serial bus, USB). Такі пристрої зазвичай реалізовані на мікросхемах FTDI232r, або ж на її аналогах, як, наприклад CH340G. В нашому випадку для підключення модулю до ПК через USB-інтерфейс ми будемо використовувати конвертер на базі FTDI232r (рис. 3), у випадку ж з ESP-12E, який ми використовуємо у вигляді плати для розробки NodeMCU1.0, використовується схема CH340G, оскільки вона вже вбудована в плату NodeMCU.

Слід зазначити, що для коректної роботи конвертер має підтримувати роботу з напругою 3,3 В, так як при напрузі 5 В є ймовірність вивести з ладу МК ESP8266. В випадку використання такого конвертеру, як зображений на рис. 3, його можна перемикати як в режим роботи на 3,3 В так і на 5 В перемикаючи відповідну перемичку на фронтальній стороні пристрою.

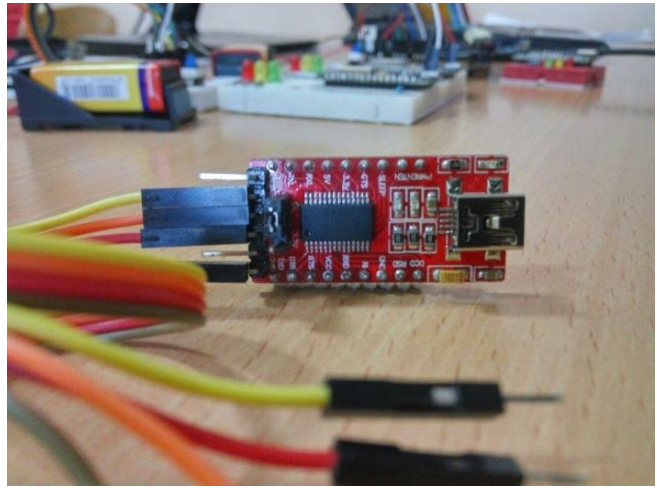

*Рис. 3. USB-UART конвертер на базі мікросхеми FTDI232r*

Далі необхідно з'єднати відповідні GPIO на конвертері та модулі. До контакту RX на модулі треба під'єднати контакт TX на конвертері (оскільки у разі, коли модуль надсилає дані — конвертер їх приймає і навпаки), а TX на модулі до RX на конверторі. Також необхідно з'єднати контакти GND між собою. У разі, якщо ESP-01 не отримує живлення від стороннього джерела — можна живити його від конвертера. Для цього необхідно з'єднати контакти VCC на платах. Наразі ми маємо змогу прочитати дані, надіслані модулем, на ПК. Для цього можна скористатися будь-якою програмою-монітором послідовних портів.

На даному етапі ми маємо під'єднаний до ПК модуль з МК ESP8266, проте на ньому не налаштовано жодного функціоналу. Для того, аби модуль виконував певні дії, необхідно описати алгоритм його дій у вигляді програми та завантажити її до пам'яті МК. Даний мікроконтролер можливо програмувати на таких мовах програмування як С та С++, Python, Lua та JavaScript (JS). При цьому слід розуміти, що дані мови не можуть надати повного контролю над модулем, оскільки це мови високого рівня, що мають високий рівень абстракції, але не дають прямого контролю над апаратною складовою. Окрім цього враховуючи досить скромні характеристики пристрою, запускати на ньому додаткові програми-інтерпретатори не є доцільним. Тому краще писати програму для модулю на мовах С або С++.

Працювати з МК в даному випадку можна, використовуючи функції зі стандартного комплекту засобів розробки (software development kit, SDK), що надається розробником модуля, або ж використовуючи бібліотеку для інтегрованого середовища розробки (integrated development environment, IDE) Arduino, або ж об'єднуючи ці способи, оскільки дана бібліотека є лише певною «обгорткою» над функціями, наданими в SDK.

У першому випадку ми маємо змогу працювати в будь-якому редакторі або IDE, використовуючи для компіляції та завантаження програми у модуль надані у SDK інструменти. Другий же варіант можливий лише при використанні Arduino IDE або ж при додаванні розширень, що додають функціонал Arduino IDE до інших середовищ розробки (наприклад, розширення VisualMicro для Microsoft Visual Studio).

У даному випадку, нами буде використовуватися середовище Arduino IDE з бібліотекою ESP8266, оскільки таке рішення є оптимальним з розрахунку «ефективність/складність». При використанні даного середовища розробки ми можемо працювати, використовуючи, як стандартні бібліотеки мов С та C++, так і використовувати бібліотеки Arduino, що спрощує написання програми.

## 4. РОЗРОБКА МЕРЕЖІ ІЗ САМООРГАНІЗАЦІЄЮ

Для проведення тестувань можливостей сенсорної ad hoc мережі, побудованої на базі стандарту IEEE 802.11 ми спробуємо побудувати власну мережу, що скрадатиметься із модулів на МК ESP8266 з самостійно розробленою мікропрограмою.

Так як найбільш широко такого роду мережі використовуються у IoT рішеннях, можемо зробити висновок, що головною функцією модулів у мережі є отримання невеликої кількості даних від пристроїв, до яких вони під'єднанні через послідовний інтерфейс, передача цих даних по каналу Wi-Fi між модулями та повернення результату батьківському пристрою через послідовний інтерфейс. Відповідно, система «модуль пристрій» має як отримувати, так і опрацьовувати отримані команди. Тому для моделювання даних умов, для зручності ми будемо використовувати плату розробника NodeMCU версії 1.0 з встановленим на неї модулем ESP-12E. Використання даної плати дозволить нам проводити налагодження мікропрограми модуля без додаткового обладнання, оскільки на платі встановлено USB інтерфейс, кнопку перезавантаження МК, вбудовано керований світлодіод, виведено 30 GPIO. Все це значно пришвидшує роботу над проектом. Можливість підключити плату напряму до комп'ютера дає змогу автоматично завантажувати код з програмного середовища, а розпаяні GPIO дозволяють під'єднати до плати світлодіоди, що будуть імітувати виконання реальних задач пристроєм, якому було надіслано команду. В подальшому ж, після отримання робочого зразка програми, проект можна буде абсолютно безболісно перенести на модуль ESP-01 та будь-яку плату з послідовним інтерфейсом, як, наприклад, Arduino Nano (рис. 4).

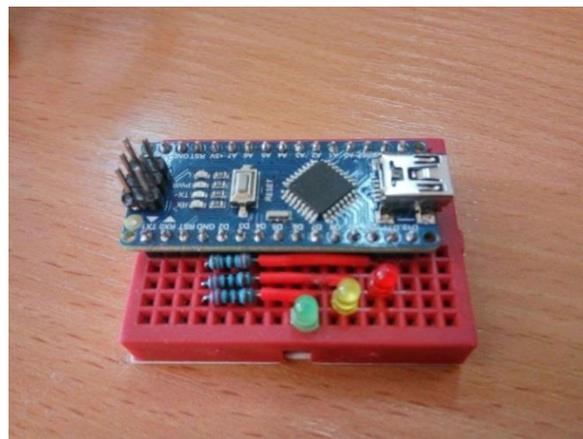

*Рис. 4. Тестовий пристрій на базі модуля Arduino Nano*

Для початку роботи нам необхідно мати принаймні дві таких плати, аби мати можливість протестувати можливість як передавати, так і приймати повідомлення. Тому для кожної з плат необхідно взяти по макетній платі, на яку ми встановимо самі плати NodeMCU, по 3 світлодіоди та з'єднаємо перемичками з GPIO 4, 5 та 15 на платі. Для зручності макетні плати можна з'єднати між собою.

В результаті маємо приблизно такий тестовий стенд, необхідний для початку розробки і виконання перших тестів, зображений на рис. 5.

Для побудови першого тестового стенду було використано такі елементи та мікросхеми:

– дві плати розробника NodeMCU 1.0;
– два з'єднувальних кабелі USB — Micro USB;
– дві макетні плати розміром 5,5×8,5 см;
– по два червоних, жовтих та зелених світлодіоди;
– два резистори на 1 кОм.

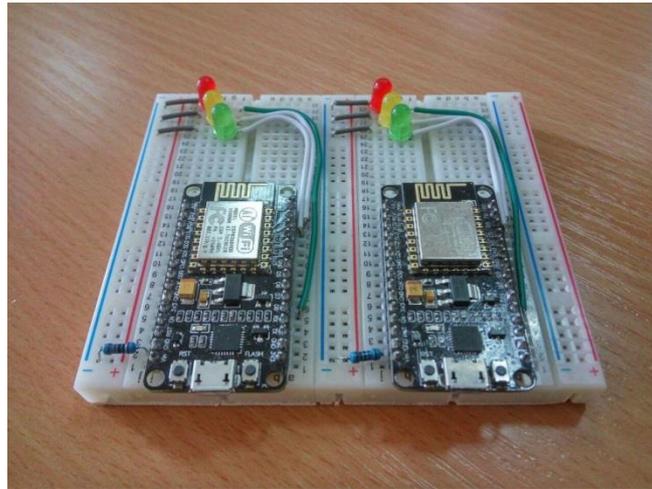

*Рис. 5. Тестовий стенд з двох модулів NodeMCU1.0*

Проте варто враховувати, що подібне рішення доступне лише для початкових етапів тестування, оскільки на платі NodeMCU використовується модуль ESP-12E, замість ESP-01, який планується використовувати з кінцевими пристроями. А також слід врахувати те, що модуль ESP-12 коштує майже удвічі дорожче за ESP-01, тому його масове використання складніше як для практичного користування, так і для подальших тестів. Тому для подальших тестів було також зібрано тестову плату з модулем ESP-01, що живиться від елементу живлення типу 6F22 (або ж «Крона»), що є зручним варіантом для живлення такого роду пристрою через свою доступність, довговічність та ємність до 500 мА·год. Схема кінцевого пристрою представлена на рис. 6, загальний вигляд модулю зображено на рис. 7.

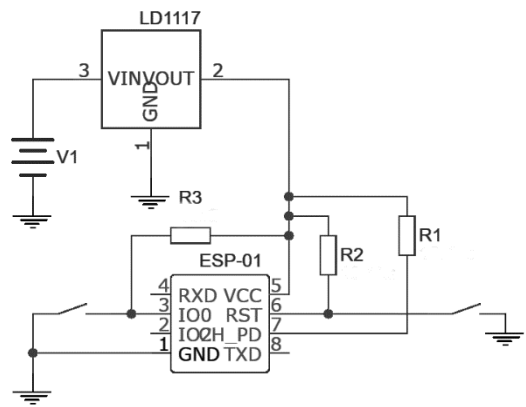

*Рис. 6. Принципова схема тестового модулю на базі ESP-01*

Єдиним недоліком рішення є напруга, яку дає дана батарея: до 9 В. При цьому ESP8266 може працювати при напругах від 3 до 3,6 В. Проте питання можна вирішити за допомогою відповідного стабілізатора напруги. У даному випадку було використано стабілізатор LD1117. Також замість нього можна використовувати й аналогічні стабілізатори, як, наприклад, AMS1117 або LM1117. В результаті було зібрано зображений на рисунку тестовий пристрій, який є повністю автономним, тому при реалізації на повноцінній платі, його можна використовувати в умовах зі ускладененням доступу, таким чином, використовуючи пристрій для отримання даних від пристроїв, що знаходяться в віддалених точках, при цьому без необхідності прямувати до них.

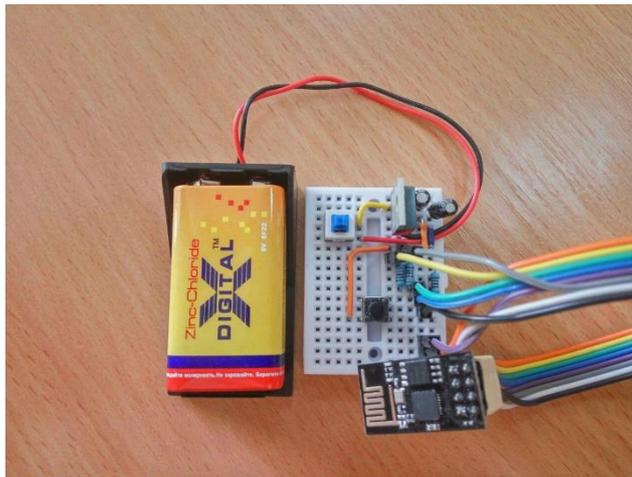

*Рис. 7. Загальний вигляд тестового модулю на базі ESP-01*

Для розгортання проекту на ПК з ОС Windows 10 необхідно встановити Arduino IDE, завантаживши інсталяційний пакет з офіційного сайту [7]. Процес встановлення середовища розробки стандартний для програм в середовищі Windows.

Після встановлення даного ПЗ необхідно встановити ESP8266 SDK та необхідні бібліотеки. Для встановлення набору розробника необхідно запустити середовище розробки, де перейти до вікна керування сторонніми бібліотеками Library Manager, де необхідно знайти та встановити пакет з назвою *ESP8266 by Simon Peter, Markus Sattler, Ivan Grokhotkov*. При цьому встановляться всі необхідні для роботи з ESP8266 бібліотеки. Але бібліотеку ESP8266WiFiMesh краще завантажити з її Git-репозиторію [8] та встановити вручну у папку *C:\Users\%username%\Arduino\libraries*, де *%username%* — ім'я вашого користувача у системі. Це варто зробити через деякі параметри у вихідних кодах даної бібліотеки, які варто буде виправити вручну. Тому одразу після встановлення вихідного коду бібліотеки до папки */libraries/*, треба одразу перейти туди та, виконати описані нижче дії.

По-перше змінюємо назву папки на іншу для уникнення конфліктів з оригінальною бібліотекою. Так, у нашому випадку папку було перейменовано на *ESP8266WiFiMesh_custom*. Далі переходимо у папку, відкриваємо за допомогою редактора файл *library.properties*, де так само змінюємо значення *ESP8266WiFiMesh* на *ESP8266WiFiMesh_custom*. Такі ж дії виконуємо у файлі *keywords*, за виключенням рядка *ESP8266WiFiMesh KEYWORD1*, його залишаємо без змін. Переходимо у папку */src/*. В ній лежать файли *ESP8266WiFiMesh.cpp* та *ESP8266WiFiMesh.h*, до назви яких також необхідно додати закінчення _custom; у файлі *ESP8266WiFiMesh_custom.cpp* знаходимо рядок:

```
#include "ESP8266WiFiMesh.h"
```

та змінюємо на:

```
#include "ESP8266WiFiMesh_custom.h"
```

На даному етапі ми внесли в файли бібліотеки всі необхідні зміни, необхідні для того, аби в подальшому мати окремо робочу версію бібліотеки, в яку можна вносити всі необхідні правки, тож тепер можемо перейти до змін, які стосуються безпосередньо функціоналу бібліотеки. В першу чергу на самому початку файлу необхідно знайти рядок:

```
#define SSID_PREFIX "Mesh_Node"
```

та замінити в ньому значення на:

```
#define SSID_PREFIX "Node"
```

Дана зміна не є суттєвою, проте завдяки їй ми скорочуємо префікс, з якого починатиметься унікальне ім'я кожного з вузлів в нашій мережі. Це в деякій мірі скоротить розмір повідомлень, а також зробить повідомлення краще сприймаються людиною.

Друга зміна є вкрай важливою, та впливає на алгоритм роботи мережі в цілому. Для кращого розуміння відмінностей стандартного та запропонованого мною рішення, нам необхідно знайти тіло методу *ESP8266WiFiMesh::attemptScan()*, у якому безпосередньо описано алгоритм пошуку потенційних точок доступу, тобто вузлів, до яких можна підключитися для передачі даних. Даний метод сканує Wi-Fi мережу на наявність в ній ТД з відповідним префіксом (в нашому випадку *Node*) і циклічно перебирає кожну таку ТД. У тілі циклу нас цікавить безпосередньо наступна частина:

```
if (index >= 0 && (target_chip_id < _chip_id)) {
    WiFi.mode(WIFI_STA);
    delay(100);
    connectToNode(current_ssid, message);
    WiFi.mode(WIFI_AP_STA);
    delay(100);
}
```

а саме (*target_chip_id < _chip_id*) в умові. Через цю умову, при переборі доступних ТД модуль звіряє свій унікальний номер (змінна *chip_id*) з унікальним номером іншого модуля (*target_chip_id*), який він передає у своєму SSID, і підключається до нього лише у тому разі, якщо номер цільового модуля менший за власний. У іншому ж випадку модуль не буде підключатися до знайденої ТД і єдиним способом передати інформацію залишиться лише очікування, коли до нього під'єднається інший модуль, у якого ідентифікатор буде меншим за ідентифікатор нашого модуля. Тоді він зможе передати повідомлення у відповіді на запит того, хто підключився. Як можна здогадатися, подібна перевірка виконана для найпростішої синхронізації між модулями, аби зв'язок між кожною парою пристроїв був однозначним. Проте, у такому разі для повноцінної двосторонньої передачі даних ми маємо генерувати запити на підключення навіть у тому разі, якщо повідомлення у черзі відсутні, для того, щоб інший модуль міг надіслати повідомлення у відповіді на запит. Така схема займає як велику кількість процесорного часу, змушуючи його виконувати зайві дії, так і завантажує мережу передачею пустих повідомлень. Тому було зроблено спробу вилучити перевірку ідентифікаторів, аби кожен з модулів міг підключитися до іншого і робив це лише у випадку, якщо у нього є повідомлення. Для цього необхідно замінити рядок:

```
if (index >= 0 && (target_chip_id < _chip_id))
```

на:

```
if (index >= 0)
```

Експериментальним шляхом було з'ясовано, що подібна зміна ніяким чином не шкодить процесу обміну даними, при цьому в певній мірі пришвидшуючи його, та оптимізувавши сам алгоритм, як було описано вище.

Залишається лише встановити бібліотеку, яка дозволить нам вираховувати контрольну суму для повідомлень. Для цих цілей було знайдено відповідний проект на GitHub. Встановити його можна так само, як і попередню бібліотеку, завантаживши її з репозиторію [8] та додавши до папки */libraries/*.

Наступним етапом роботи стала розробка мікропрограми для системи, що була зібрана. Вже на етапі розробки архітектури програми з'явились деякі складності. Так як повноцінні ad hoc мережі зазвичай працюють на багатофункціональному обладнанні призначеному безпосередньо для побудови мереж (Wi-Fi точки доступу та маршрутизатори), яке має значні обчислювальні можливості та здатність до багатопоточної обробки даних, постала проблема як реалізувати ті ж самі функції на мікроконтролері, що значно поступається своїми технічними характеристиками. Для вирішення цього питання було розглянуто ряд рішень з відкритим вихідним кодом, зокрема вище згадана бібліотека ESP8266WiFiMesh. В результаті вдалося з'ясувати, що всі подібні рішення реалізовані на почерговій зміні режимів роботи модулю, як це вказано на рис. 8.

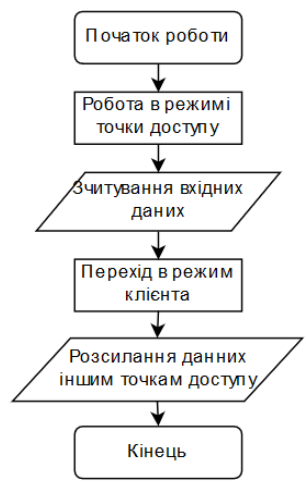

*Рис. 8. Загальний алгоритм роботи Mesh-мережі на МК*

Алгоритм обміну було майже повністю реалізовано за допомогою бібліотеки ESP8266Mesh, а отже і режим роботи модуля керується алгоритмом зображеним вище, проте з невеликою зміною, яка була описана вище. В кожному етапі циклу мікропрограма прослуховує вхідні підключення, у разі, якщо якийсь з модулів підключився до нього — обробляє вхідне повідомлення та надсилає назад відповідь. Наступним кроком перевіряє повідомлення в черзі та у разі їх наявності, перемикаючись в режим клієнта, надсилає перше повідомлення з черги суміжним модулям. Далі процедура повторюється.

Для зберігання повідомлень, що повинні бути надіслані, було створено дві спеціальні структури. Перша структура з назвою *msg* призначена для зберігання і обробки безпосередньо самих повідомлень. Дана структура дозволяє зберігати повідомлення, та його мета-дані окремо, але в єдиному об'єкті, що в свою чергу полегшує роботу з цими даними та пришвидшує доступ до них. Структура *msg* складається з таких елементів:

– *id* — унікальний номер повідомлення, за яким воно ідентифікується;

– *from* — ім'я МК, який надіслав повідомлення;

– *to* — ім'я МК, якому адресовано повідомлення;

– *media* — ім'я МК, від якого прийшло повідомлення;

– *ttl* — «час життя пакету» (time to live, TTL) повідомлення. Вказує скільки разів можна переслати повідомлення перш ніж воно втратить актуальність;

– *body* — тіло повідомлення. Тобто безпосередньо сам текст, який необхідно переслати;

– *crc* — контрольна сума пакета (CRC).

Окрім цього структура має в собі чотири функції для роботи з повідомленнями:

– *newMessage(String addressee, String message)* заповнює усі елементи структури з усіма мета-даними на основі самого повідомлення, що передається у змінній *message* та ім'я вузла, якому це повідомлення адресовано (змінна *addressee*);

– *show()* відображає дані збережені у змінній типу *msg* у зручночитаємому вигляді;

– *string2msg(String message)* конвертує передане повідомлення з формату *String* (у якому воно пересилається між МК) у формат *msg* для подальшої обробки та зберігання;

– *msg2string()* виконує зворотню функцію, а саме формує рядок типу *String* з повідомлення для подальшого пересилання.

Друга структура має назву *storage*. Вона призначена безпосередньо для зберігання повідомлень та їх обробки. В ній формується черга типу FIFO (first input, first output), яка дозволяє опрацьовувати всі повідомлення по мірі їх надходження. Вона складається з наступних змінних:

– *message* — тіло повідомлення у форматі *msg*;

– *timer* — лічильник, необхідний для черги *sended*, що вказує скільки циклів буде зберігатися повідомлення;

– *size* — лічильник, який зберігає кількість повідомлень у черзі;

– *sent* — лічильник, що рахує скільки разів було надіслане перше повідомлення в черзі. Необхідний для подальшого перенесення повідомлення з однієї черги до іншої;

– *\*first* — вказівник, що вказує на перший елемент в черзі для того, аби знати звідки починати діставати елементи;

– *\*last* — вказівник, що вказує на останній елемент в черзі. Нові елементи додаються після цього елемента;

– *next* — вказівник, що вказує на наступний елемент. Зв'язок між повідомленнями в черзі реалізовано через нього.

Та таких функцій:

– *add()* — додає нове повідомлення в кінець черги;

– *select1st()* — повертає значення першого повідомлення в черзі;

– *destroy()* — видаляє перше повідомлення в черзі;

– *showAll()* — відображає усі повідомлення у зручному для сприйняття вигляді;

– *search()* — шукає у черзі повідомлення за його ідентифікатором та повертає *true* або *false* в залежності від результату;

– *timerDec()* — зменшує параметр *timer* кожного повідомлення на одиницю.

Так як розмір черги може варіюватися в залежності від завантаженості вузла та швидкості розсилання повідомлень, збереження даних виконується у динамічну пам'ять. У зв'язку з цим виникли деякі складності під час написання даної частини програми. Через допущення невеликої помилки у роботі з вказівниками, програма зазнавала фатальної помилки та МК аварійно перезавантажувався. Як вдалося виявити, проблема була пов'язана з вказівником, що з'єднував повідомлення в черзі між собою, через що програма посилалася на вже зайняту ділянку пам'яті. Після виправлення помилки було отримано перший робочий прототип програми, завдяки якому можна було реалізувати передачу даних між двома тестовими модулями NodeMCU1.0.

Проте постало наступне питання: як захистити мережу від утворення петлі, коли повідомлення, що було відправлене вузлом повертається до нього, він його знову додає до черги і знову відправляє. Для вирішення даної проблеми необхідно було десь зберігати дані про відправленні повідомлення, аби контролер «пам'ятав», що він вже відправив і більше цих повідомлень не отримував. Для цієї мети до програми було додано другу чергу типу *storage*, у яку на певний час переносяться усі відправленні повідомлення. Таким чином почали зберігатися у двох змінних типу *storage*: to_send та sended для повідомлень, що необхідно розіслати та них, що вже розіслано відповідно.

У черзі *to_send* перевіряється змінна *sent*, яка інкрементується при отриманні підтвердження передачі, та якщо повідомлення було успішно комусь надіслано, спочатку додає його до черги *sended*, після чого видаляється з черги *to_send*. У консоль при цьому виводиться службове повідомлення з вказанням кількості вузлів, яким було надіслано повідомлення. Після цього змінна *sent* обнуляється.

В останньому рядку приведеного фрагменту коду реалізована регуляція повідомлень у черзі *sended*, через виклик функції *timerDec()*, тіло якої наведено нижче:

```
void storage::timerDec() {
    storage *handler = first;
    if (handler != NULL) {
        if (handler->timer <= 0)
        destroy();
        while(handler != NULL) {
            handler->timer--;
            handler = handler->next;
        }
    }
}
```

Дана функція шукає перший елемент в черзі, та орієнтуючись на нього починає перебирати всі елементи, декрементуючи змінну *timer* в кожному повідомленні. Якщо ж у повідомленні дана змінна менша за 0 — викликається функція *destroy*, що видаляє перше повідомлення з пам'яті та підставляє адресу наступного вказівника *\*first*. Реалізація функції *destroy* виглядає наступним чином:

```
void storage::destroy() {
    storage *handler = first;
    if (handler != NULL) {
        first = handler->next;
        size--;
        delete handler;
    }
}
```

Оскільки у даному модулі є лише близько 40 КБ вільної динамічної пам'яті, з якою може працювати МК, одразу постає питання, скільки повідомлень може зберігати такий модуль. Для тестування даного питання було написано тестову функцію *test()*, яка просто генерувала тестове повідомлення та додавала його до черги. Описані вище функції для автоматичного очищення черги при цьому були вимкнені. Функція *test()* описана нижче:

```
void test() {
    msg test_msg;
    String test_s;
    test_s = "MSG:01N02|Node1/Node3/Node2|8|TESTTESTTEST|0000";
    test_msg.string2msg(test_s);
    sended.add(test_msg);
}
```

Завдяки даному способу було виявлено, що загальна кількість повідомлень з обох черг не повинна перевищувати 198-ми повідомлень, при довжині тексту в повідомленні до 32-х символів. Саме на таку кількість вистачає приблизно 40 КБ динамічної пам'яті пристрою. Тому до функції *add()* була додана перевірка змінної *size*, яка зберігає кількість повідомлень у черзі. Якщо ж вона більша за 100 — повідомлення до черги додане не буде, а в консоль виведеться повідомлення про переповнення черги.

МК отримує повідомлення за допомогою виклику методу *acceptRequest()* класу ESP8266WiFiMesh, який циклічно викликається у функції *loop()*. Після того, як повідомлення було отримане, воно передається функції *ManageRequest(String request)* у вигляді стрічкової змінної типу *String* для подальшої обробки. Саме повідомлення має такий вигляд:

```
MSG:xxNyyy|NodeYYY/NodeYYY/NodeZZZ|N|text|crc
```

де *MSG* — мітка, що вказує на те, що дане повідомлення повинно бути опрацьоване за вказаним алгоритмом; *xxNyyy* — ідентифікатор повідомлення, де *xx* — порядковий номер повідомлення, що надіслав певний модуль від початку своєї роботи, а *xxx* — унікальний номер даного модуля. Його було додано для уникнення ситуації, при якій від різних модулів може прийти повідомлення з однаковим порядковим номером й відповідно однаковим ідентифікатором. Літера *N* у даному випадку виступає у вигляді роздільного символу, що ділить дві частини ідентифікатора; NodeYYY/NodeYYY/NodeZZZ — ідентифікатор модуля відправника, модуля-посередника та модуля адресата відповідно. *YYY*, *ZZZ* та *XXX* у даному випадку — унікальні номери кожного з модулів. Якщо повідомлення прийшло від модуля без участі посередників — поля відправника та посередника збігатимуться; *N* — TTL повідомлення; *text* — безпосередньо команда, яку необхідно переслати; *crc* — контрольна сума повідомлення, розрахована за допомогою сторонньої бібліотеки CRC32 для Arduino. Кожна з частин розділяється за допомогою спеціального символу «|» («конвеєр» або ж "pipeline") для зручності подальшого розбиття на складові.

В першу чергу функція *ManageRequest()* перевіряє початок повідомлення, де знаходиться мітка, що вказує тип повідомлення. Всього було закладено два типи повідомлень, що мають такі мітки: *MSG* — основний тип повідомлень, призначений для передачі безпосередньо інформації та *ST* — мітка для службових повідомлень, що використовується при надсиланні відповідей.

Далі, в залежності від умов, у відповідь на отримане повідомлення може бути надіслана одна з таких відповідей:

– *ST:BROKEN* — контрольна сума повідомлення не збігається. При отриманні такої відповіді модуль спробує надіслати повідомлення ще раз. Контрольна сума формується щоразу при передачі повідомлення від модуля до модуля та перевіряється при отриманні. Формується вона з рядку, що складається з ідентифікаторів відправника, отримувача та посередника, ідентифікатора повідомлення та тіла повідомлення. При вирахуванні контрольної суми використовується алгоритм CRC-32 стандарту IEEE;

– *ST:OK* — повідомлення отримане та опрацьоване. У разі отримання такої відповіді, модуль, що надсилав його, видалить повідомлення з черги *to_send* та перенесе до черги *sended*;

– *ST:EXPIRED* — TTL повідомлення рівний або менший за нуль. Наразі не відрізняється від ST:OK. Просто означає, що модуль-отримувач отримав, проте ніяк не опрацював повідомлення через його неактуальність;

– *ST:DUBLICATE* — означає, що у сховищі модуля-отримувача вже є подібне повідомлення, тому він не буде опрацьовувати його ще раз. Перевірка відбувається по ідентифікатору повідомлення, за допомогою функції *search(String mes_id)*, якій передається ідентифікатор повідомлення. Вона послідовно перебирає усю чергу та повертає істину, якщо у черзі є повідомлення з відповідним номером. У іншому разі повертається хиба.

Як видно з описаного функціоналу, на кожне повідомлення зазначеного зразка, формується певна відповідь, від якої залежить подальша обробка повідомлення. Це забезпечує надійність передачі та гарантує, що повідомлення дійде до адресата принаймні одним з випадково утворених шляхів.

Надсилаються повідомлення за допомогою функції *getMessage()*, яка перевіряє, чи є повідомлення у черзі, так у разі їх наявності переводить його з формату *msg* до формату *String* та викликає бібліотечну функцію *attemptScan()*, яка намагається передати повідомлення, що було їй передано всім ТД, які вона знайде. При цьому на NodeMCU реалізовано спалахування вбудованого світлодіоду під час передачі. На ESP-01 такий функціонал відсутній.

Так як головною метою використання нашого рішення є обмін даними між іншими пристроями та побудова мереж IoT, можна зробити висновок, що модулі повинні виступати не як самостійні пристрої, а лише як додаток до працюючих пристроїв, що має надати їм можливість до комунікації. Тому для повноцінного тестування функціоналу розроблюваного рішення необхідно, аби вони могли передавати дані іншим пристроям через послідовний інтерфейс. Тому у випадку с ESP-01 буде використано плату для розробки Arduino Nano (див. рис. 4) з під'єднаними до неї світлодіодами, що відображатимуть роботу плати. Для самої плати було написано невелику програму, що при отриманні даних від ESP-01 обробляє їх та запалює відповідні світлодіоди. А також плата час від часу генерує тестові повідомлення для одного з вузлів мережі задля створення навантаження на мережу і її подальшого тестування.

У разі ж використання плати NodeMCU1.0, велика кількість попередньо виведених з неї GPIO дає змогу використовувати для тестів саму плату. Тому до цих плат також під'єднані світлодіоди, а підпрограма для Arduino Nano буде додана і до загального коду для ESP8266.

До плати під'єднано три світлодіоди. Їх запалювання керується командою *set_led* за параметром у вигляді цифри від 0 до 7, де кожній цифрі відповідає своя комбінація. Комбінації розподілено з використанням двійкової системи числення, де ввімкнений режим дорівнює одиниці. Для кращого розуміння звернемося до табл. 2 (де «0» — вимкнено, а «1» — увімкнено):

*Таблиця 2*

**Пояснення кодів тестової команди set_led**

| Десяткове число / Світлодіод | 0 | 1 | 2 | 3 | 4 | 5 | 6 | 7 |
|---|---|---|---|---|---|---|---|---|
| Червоний | 0 | 1 | 0 | 1 | 0 | 1 | 0 | 1 |
| Жовтий | 0 | 0 | 1 | 1 | 0 | 0 | 1 | 1 |
| Зелений | 0 | 0 | 0 | 0 | 1 | 1 | 1 | 1 |

Відповідно, наприклад, команда *set_led* = 3 запалить червоний та жовтий світлодіоди, та залишить зелений вимкненим, бо 3 в двійковій системі числення буде 110, де перші дві одиниці відповідають запаленим світлодіодам.

Окрім цього в модулі реалізовано кілька службових команд для керування самим модулем, що мають на меті допомогти користувачу у відладці та тестуванні пристрою. Серед них:

– *get_id* — виводить ідентифікатор пристрою до консолі;

– *get_sended* — відображає у консолі все повідомлення у черзі *sended* та їх кількість;

– *get_send* — відображає у консолі всі повідомлення у черзі *to_send* та їх кількість.

Дані команди не несуть практичної користі беспосередньо при користуванні, проте значно спрощують роботу з модулями під час дослідження.

Головним типом даних, що находять до модуля через послідовний інтерфейс є команда, яка додає нове повідомлення до черги і в результаті надсилає його. Вона має такий вигляд:

```
Node<унікальний номер модулю>@<команда>
```

де, в першій частині команди вказується адресат, кому необхідно надіслати це повідомлення, а в другій вказується тест повідомлення. Також можливо звернутися і до модулю, до якого ми звертаємося через послідовний інтерфейс, просто вказавши слово *this* замість ідентифікатора. Тоді повідомлення не додаватиметься до черги, а одразу буде оброблено. Робота через даний інтерфейс реалізована через функцію *getSerialData()*, що викликається у функції *loop()* та таким чином прослуховує введені через консоль дані. Після їх запису у змінну, модуль перевіряє, чи підходять вони під шаблон, вказаний вище. Якщо повідомлення некоректне — видається попередження, та надісланий текст надалі не обробляється. Якщо ж повідомлення коректне, далі перевіряється перша частина команди, якщо там вказано ідентифікатор, створюється і передається до черги повідомлення. Якщо ж там стоїть слово *this* — текст команди передається до функції *manageCommand()*, де далі перевіряється на відповідність командам, які ми розглянути вище.

## 5. ЕКСПЕРИМЕНТАЛЬНО ОЦІНКА САМООРГАНІЗАЦІЇ СИСТЕМИ

Експериментальна установка складається з робочої моделі мережі, в яку входять три вузли (дві плати NodeMCU з тестового стенду та тестовий пристрій з модулем ESP-01).

Першим тестом стала безпосередньо передача даних між вузлами та обробка отриманих повідомлень вузлом. З цією метою було відправлено два тестових повідомлення з вузла Node14754480, яке адресувалося вузлу Node52126. У тестовому повідомленні було відправлено команду, яка мала запалити світлодіоди на модулі Node52126. При цьому модуль отримав повідомлення двома шляхами: напряму від модуля відправника та через «посередника» Node1592748, який також отримав повідомлення від Node14754480 та переслав його далі. Друге повідомлення було відкинуючие як дублююче. Система спрацювала коректно. Нижче наводиться виведені до консолі повідомлення від модулів Node14754480 та Node52126.

Команди виведені першим модулем:

```
Node52126@set_led=6
<[Will send:] Node14754480 -> Node14754480 -> Node52126 (8) set_led=6
>[To 2  sent]
>[DUBLICATE:] Node14754480 -> Node1592748 -> Node52126 (7) set_led=6
```

З нього видно, що після введення команди «Node52126@set_led=6» до консолі модуля, він згенерував повідомлення та додав його до черги *to_send*, після чого спробував його розіслати суміжним модулям. В результаті було виведено повідомлення «>[To 2 sent]», що означає, що модулю вдалося передати повідомлення двом іншим модулям. При цьому модуль Node52126 отримав повідомлення, та так як воно адресувалося йому, він обробив команду та не надсилав повідомлення далі. А модуль Node1592748, отримавши повідомлення, розіслав його іншим. Тому було вдруге надіслано модулю-адресату та також повернулося першому модулю, проте було відкинуте, оскільки таке ж повідомлення вже зберігається у черзі *sended*.

Розглянемо виведення до консолі другого модулю (Node52126):

```
>[MyMessage:] Node14754480 -> Node1592748 -> Node52126 (7) set_led=6
*Yellow and Green ON
>[DUBLICATE:] Node14754480 -> Node14754480 -> Node52126 (8) set_led=6
```

Модуль прослуховував вхідні запити, та отримавши повідомлення, адресоване йому, прочитав та виконав команду. Про це свідчить виведення команди «*Yellow and Green ON». Слід зазначити, що швидше він зміг отримати повідомлення від проміжного модуля Node1592748, ніж напряму від модуля-відправника. Про це свідчить рядок, що відображає шлях, який пройшло повідомлення у даному разі: Node14754480 -> Node1592748 -> Node52126 (7). При цьому цифра вказана у дужках відображає TTL повідомлення. Оскільки пакет пройшов через один додатковий вузол, його значення було зменшено на одиницю (за замовчанням значення TTL = 8).

Для підтвердження працездатності системи тест було повторено. З того ж модулю (Node14754480) було відправлене ще одне повідомлення. Вивід даного модулю:

```
Node52126@set_led=1
<[Will send:] Node14754480 -> Node14754480 -> Node52126 (8) set_led=1
>[To 1 sent]
```

В консоль було введено команду «Node52126@set_led=1», після чого модуль знову згенерував повідомлення та додав його до черги для відправлення. Але на цей раз з відповіді «>[To 1 sent]» бачимо, що повідомлення вдалося передати лише одному модулю. Подивимося при цьому виведення модулю Node52126:

```
>[MyMessage:] Node14754480 -> Node1592748 -> Node52126 (7) set_led=1
*Red ON
```

Повідомлення в даному випадку було отримане напряму від модуля-відправника та опрацьоване. Було запалено червоний світлодіод. Повідомлення дійшло до адресата, подальше розсилання не потребується. Система перейшла в режим прослуховування. Тест було пройдено успішно.

При цьому було приблизно виміряно час передачі повідомлення. В залежності від умов, передача даних у даній мережі може займати від 1 до 5 с, в залежності від шляху повідомлення та швидкості реакції модулів.

Для тестування також було задіяно тестовий пристрій на базі плати Arduino Nano, котрому модуль повинен передавати повідомлення на обробку після чого світлодіоди повинні запалюватися на даному пристрої. Дана функція працює так само як і при обробці команд зі світлодіодами на тестових модулях з NodeMCU. При цьому плата Arduino також раз у 5 с генерує повідомлення для інших модулів і передає його модулю ESP-01, аби він його відправив далі. При тестуванні було виявлено проблему з некоректністю зчитування повідомлень модулем. Причиною цьому стали спеціальні символи, які надавала плата при передачі. Після усунення даної проблеми повідомлення почали розсилатися коректно, а плата отримала можливість комунікації з іншими пристроями.

## 6. ВИСНОВКИ ТА ПЕРСПЕКТИВИ ПОДАЛЬШИХ ДОСЛІДЖЕНЬ

На основі розробленого експериментального макету і результатів дослідів було розроблено та створено власну реалізацію сенсорної ad hoc мережі на базі мікроконтролера ESP8266. Для цього було використано сторонні відкриті рішення у поєднанні з власними розробками. Використовуючи мережу побудовану на нашій платформі, було проведено ряд тестів, що дозволили краще зрозуміти її слабкі місця, аби в подальшому звернути на них увагу та продумати шляхи їх виправлення. Отримані результати можна ефективно використати для подальших робіт у даному напрямку, що дозволить удосконалити отримане рішення та в подальшому застосувати їх у реальних платформах, побудованих як на невеликих контролерах так і на повноцінних мережевих пристроях, що наразі є досить актуальним. Тож в подальшому, після певних доопрацювань дана система може бути використана в реальних умовах для утворення системи розумного дому, отримання інформації з певних датчиків.

## СПИСОК ВИКОРИСТАНИХ ДЖЕРЕЛ

**Maksym A. Vladymyrenko**
PhD student
Borys Grinchenko Kyiv University, Kyiv, Ukraine
OrcID: 0000-0001-8852-7939
*reinekekun@gmail.com*

**Volodymyr Yu. Sokolov**
MSc, senior lecturer
Borys Grinchenko Kyiv University, Kyiv, Ukraine
OrcID: 0000-0002-9349-7946
*vladimir.y.sokolov@gmail.com*

**Volodymyr M. Astapenya**
PhD, associate professor
Borys Grinchenko Kyiv University, Kyiv, Ukraine
OrcID: 0000-0003-0124-216X
*v.astapenia@kubg.edu.ua*

# RESEARCH OF STABILITY
# IN AD HOC SELF-ORGANIZED WIRELESS NETWORKS

**Abstract.** To date, there is a need for the development of efficient data and device exchange protocols that this exchange will provide, since standard protocols used in traditional networks can not fully meet the needs of a new type of network. The article describes the process of development and implementation of a full-scale model of noise-resistant and sensor network breaks. The stability of this network is achieved by building a distributed network, in which all nodes send messages to all available nodes. Wireless mobile peer-to-peer network (MANET) can be configured automatically, so the nodes in it can move freely. Wireless networks do not have the complexity of infrastructure and management, which allows devices to create and join on-the-go networks - anywhere, anytime. In this paper, the theoretical part of the functioning of such networks and the field of their use is considered. After that, an initial analysis of the available equipment used for constructing such hardware solutions was conducted. The software for developing such solutions is considered in detail, as well as examples of finished models that implement the investigated functional. After that, several variants of the model of network nodes, as well as the test device for creating a payload on the network, are collected. For this purpose, third-party open solutions were used in conjunction with their own developments. The system received a series of tests that made it possible to understand the weak and strong points of such a network and draw conclusions for the further development of the project and the creation of an improved working prototype. The article presents the basic electrical circuits of devices, the list of used equipment and software used and the photographic material of prototypes of the created system. This system can be used in real conditions to create a smart home system, obtain information from certain IoT sensors.

**Keywords:** sensory network; self-organization network; data exchange; microcontroller; Wi-Fi; fault tolerance; IoT; wireless sensor.

.